\documentclass{article}

\usepackage{arxiv}
\usepackage[utf8]{inputenc} 
\usepackage[T1]{fontenc}    
\usepackage{hyperref}       
\usepackage{url}            
\usepackage{booktabs}       
\usepackage{amsfonts}       
\usepackage{nicefrac}       
\usepackage{microtype}      
\usepackage{lipsum}         
\usepackage{graphicx}
\usepackage[numbers]{natbib} 
\usepackage{doi}
\usepackage{multirow}       
\usepackage{amsmath}

\setcitestyle{numbers}

\title{A Study on B-Cell Epitope Prediction Based on QSVM and VQC}

\date{April 16, 2025}

\author{
	Chi-Chuan Hwang\thanks{
		https://orcid.org/0000-0002-3407-9832 \\
		https://researchoutput.ncku.edu.tw/en/persons/chi-chuan-hwang.} \\
	Department of Engineering Science\\
	National Cheng Kung University\\
	\texttt{chchwang@ncku.edu.tw} \\
	\And
	Yi-Ang Hong \\
	Department of Engineering Science\\
	National Cheng Kung University\\
}

\renewcommand{\shorttitle}{\textit{B-CELL EPITOPE PREDICTION BASED ON QSVM AND VQC}}

\hypersetup{
pdftitle={A Study on B-Cell Epitope Prediction Based on QSVM and VQC},
pdfsubject={q-bio.NC, q-bio.QM},
pdfauthor={Chi-Chuan Hwang, Yi-Ang Hong},
pdfkeywords={B-cell epitope prediction, Quantum Support Vector Machine (QSVM), Variational Quantum Classifier (VQC), Quantum machine learning, Bioinformatics},
}

\begin{document}
\maketitle

\begin{abstract}
	This study explores the application of quantum computing in B-cell epitope prediction, demonstrating the potential and prospects of quantum machine learning in bioinformatics through two approaches: Quantum Support Vector Machine (QSVM) and Variational Quantum Classifier (VQC). B-cell epitope prediction is a critical task in immunological research, with profound implications for vaccine design, diagnostics, and therapeutic development. The study reviews the historical development of B-cell epitope prediction methods, from early physicochemical property-based approaches to subsequent machine learning techniques, and highlights the computational efficiency limitations of traditional methods as data scale and complexity increase. Leveraging quantum entanglement and superposition, quantum computing offers a novel pathway to address these challenges.\\
    In the study, QSVM utilizes quantum kernel functions to map data into a high-dimensional feature space, solving an optimization problem to distinguish epitopes from non-epitopes. In contrast, VQC employs a variational approach, constructing a classifier through parameterized quantum circuits and feature mapping, estimating empirical probability distributions based on quantum measurement outcomes for class prediction. Experimental results show that QSVM and VQC achieved accuracies of 70\% and 73\%, respectively, demonstrating competitiveness with traditional methods. Notably, VQC slightly outperformed in accuracy, while QSVM excelled in Matthews correlation coefficient, indicating its advantage in balancing positive and negative classes. Compared to traditional methods from 2022 and 2024, quantum approaches performed comparably or even surpassed in certain metrics, highlighting the potential of quantum computing to enhance prediction performance.\\
    However, this study also highlights challenges in applying quantum methods. Quantum circuits demand significant computational resources, and current quantum hardware limitations impact the scalability of these approaches. Additionally, VQC shows certain limitations when handling imbalanced datasets, potentially leading to prediction biases. Metrics such as the area under the ROC curve and Matthews correlation coefficient indicate that quantum methods still have room for improvement in distinguishing epitopes from non-epitopes. The study suggests that increasing the number of measurements could reduce statistical noise, though this must be balanced against computational costs. Furthermore, exploring cost-sensitive learning or hybrid quantum-classical approaches may offer effective solutions to address current challenges.\\
    Overall, this study demonstrates the feasibility and advantages of QSVM and VQC in B-cell epitope prediction, underscoring the potential of quantum computing in bioinformatics. Despite challenges related to computational resources and optimization techniques, with ongoing advancements in quantum hardware and algorithms, quantum computing is poised to provide more accurate and efficient solutions for immunology and related fields in the future, driving further progress in these domains.
\end{abstract}


\section{Introduction}
B-cell epitope prediction is a critical task in immunological research, with profound implications for vaccine design, diagnostic tool development, and the formulation of therapeutic strategies. B-cell epitopes are specific regions on an antigen recognized by B-cell receptors or antibodies, capable of triggering an immune response. Accurate identification of these epitopes is essential for developing effective vaccines, particularly in the context of emerging infectious diseases such as COVID-19. The global pandemic caused by the SARS-CoV-2 virus has highlighted the urgent need for rapid and precise vaccine design. However, traditional experimental methods for epitope identification are time-consuming and costly, making computational prediction methods an attractive alternative. Nevertheless, existing computational tools for B-cell epitope prediction remain limited in performance, often facing challenges in accuracy and generalization.\\

Traditional machine learning techniques, such as Support Vector Machines (SVM), Recurrent Neural Networks (RNN), and Hidden Markov Models (HMM), have been widely applied to this problem. For instance, early methods like BcePred utilized physicochemical properties of amino acids for prediction, while BepiPred combined HMM to model sequence patterns. More advanced methods, such as ABCpred, employed RNN, and SVMTriP leveraged SVM with tripeptide similarity for prediction. However, according to a 2021 study by Galanis et al., even the best traditional methods (e.g., BepiPred) achieved a Matthews Correlation Coefficient (MCC) of only approximately 0.0778, indicating significant limitations in addressing the complexity of biological data.\\

These limitations stem from the high-dimensional feature space and nonlinear relationships in biological data, whereas quantum computing technology holds promise to address these challenges through its unique advantages. Quantum computing leverages the properties of quantum superposition and entanglement, enabling parallel processing of vast amounts of data and efficient handling of high-dimensional spaces. Quantum Machine Learning (QML) algorithms, which combine quantum computing with machine learning techniques, have shown potential in enhancing classification task performance, particularly in domains where data complexity exceeds the capabilities of traditional computers.\\

This paper focuses on the application and comparison of two major QML algorithms—Quantum Support Vector Machine (QSVM) and Variational Quantum Classifier (VQC)—in B-cell epitope prediction. QSVM extends the functionality of traditional SVM by mapping data to a high-dimensional quantum feature space through quantum kernel functions, potentially improving data separability. In contrast, VQC utilizes parameterized quantum circuits optimized via variational methods to establish classification boundaries, offering a flexible quantum-enhanced learning framework. By comparing these two quantum methods with traditional SVM, this study aims to evaluate whether quantum computing can provide superior performance in epitope prediction, particularly in scenarios with limited data or high-dimensional feature spaces.\\

The motivation for this study is twofold. First, the COVID-19 pandemic has underscored the necessity for rapid vaccine development, and precise epitope prediction can significantly accelerate this process. Second, quantum computing technology is rapidly advancing, with platforms like IBM’s Qiskit providing accessible tools for QML research. The urgency of biological demands combined with technological innovation presents a unique opportunity to explore quantum methods for addressing complex bioinformatics problems.\\

Prior studies have laid the foundation for this paper. For example, Havliček et al. demonstrated the potential of QSVM in classifying synthetic datasets, while Stoudenmire and Schwab applied tensor networks related to quantum circuits to supervised learning tasks. In the field of bioinformatics, quantum computing has been used for tasks such as protein folding and genomic sequence analysis, but its application in epitope prediction remains underexplored. This study aims to fill this research gap through a systematic comparison on real B-cell epitope datasets.\\

\section{Literature Review}
\label{sec:headings}

B-cell epitopes are critical regions of antigens that can be recognized by B cells and trigger an immune response. Accurate prediction of B-cell epitopes is of great significance for accelerating vaccine design and reducing research and development costs, particularly in the era of the COVID-19 pandemic, where this need has become especially urgent. Since Hopp and Woods developed the first amino acid sequence-based epitope prediction method in 1981 \cite{hopp1981}, this field has gradually garnered increasing attention. Early methods primarily relied on the physicochemical properties of amino acids. In 2004, Saha and Raghava introduced BcePred \cite{saha2004}, which predicts epitopes based on physicochemical propensity scales such as hydrophilicity and antigenicity. In 2006, Larsen et al. developed BepiPred \cite{larsen2006}, the first prediction method to employ a Hidden Markov Model (HMM). In the same year, Saha and Raghava proposed ABCpred \cite{saha2006}, which utilized Recurrent Neural Networks (RNNs) and improved prediction performance by training networks with varying sizes and hidden units. In 2008, Ansari and Raghava introduced PEPITO \cite{ansari2008}, enhancing the prediction of discontinuous epitopes through multiple distance thresholds and hemispheric exposure features. In 2009, Sweredoski and Baldi proposed COBEpro \cite{sweredoski2009}, which adopted a Support Vector Machine (SVM) model to assign epitope propensity scores to peptide sequence fragments. In 2012, Yao et al. developed SVMTriP \cite{yao2012}, also based on SVM, combining tripeptide similarity and propensity features for prediction. In 2013, Singh et al. introduced LBtope \cite{singh2013}, another sequence-based prediction tool. In 2015, Saravanan and Gautham developed LBEEP \cite{saravanan2015}, incorporating a new amino acid feature descriptor—Dipeptide Deviation from Expected Mean (DDE)—and combining SVM with AdaBoost-Random Forest techniques. In 2017, Jespersen et al. proposed BepiPred-2.0 \cite{jespersen2017}, which integrated sequence and conformational features and used a Random Forest model to further optimize prediction performance. In 2021, Collatz et al. reviewed the performance of various linear B-cell epitope prediction tools accessible via command-line interfaces \cite{collatz2021}. In 2023, Liu et al. developed a structure-based prediction model \cite{liu2023}, integrating local and global features and employing deep learning techniques for training.\\

In 2021, Collatz et al. \cite{collatz2021} conducted a performance comparison of multiple prediction methods. The results showed that BepiPred achieved the highest Matthews Correlation Coefficient (MCC) value (0.0778) among per-peptide methods, followed by the Consensus\_NoLBEEP algorithm (MCC = 0.0721). LBEEP performed the worst (MCC = -0.0103), while BcePred and SVMTriP had MCC values of 0.0251 and 0.0290, respectively. Their proposed Consensus\_ALL method performed best in terms of accuracy, reaching 55.59\%. In the per-residue method evaluation, their method slightly outperformed BepiPred with an MCC of 0.0489 compared to 0.0488, with accuracies of 53.04\% and 52.88\%, respectively.\\

The results described above were all achieved using classical computers. However, whether existing quantum computing platforms can be leveraged to study this topic remains unexplored, with no appropriate suggestions or research available. The purpose of this paper is to investigate this issue using the methodologies of Quantum Support Vector Machine (QSVM) and Variational Quantum Classifier (VQC). By utilizing QSVM, high-dimensional data can be efficiently classified to identify decision boundaries, while VQC employs quantum variational algorithms to transfer quantum computation results to classical computers for optimization, iterating to achieve the best outcomes. In the following sections, we will first introduce the QSVM methodology, followed by the VQC methodology. Finally, we will comprehensively compare our computational results with those obtained using classical computers by previous researchers and propose directions for future sustainable development.\\

\section{Introduction to QSVM Methodology}
Both QSVM and VSM rely on inner products to measure the similarity between data points. QSVM achieves this through a quantum feature mapping \( \phi(x) \), while VSM uses classical TF-IDF representations. Classical SVM methods are well-documented in numerous books and papers, so we will not elaborate on them here.\\

The input dataset is mapped to a higher-dimensional feature space as follows:\\
\begin{equation}
	K\left(x_i, x_j\right) = \left\langle \phi\left(x_i\right), \phi\left(x_j\right) \right\rangle 
\end{equation}

Where \( K \) is the kernel function, \( x_i \) and \( x_j \) are \( n \)-dimensional input data points, and \(\phi\) maps the \( n \)-dimensional data to an \( m \)-dimensional space. The right-hand side of equation (1) represents the inner product of the \(\phi\) function in the \( m \)-dimensional space.

For convenience, in the training dataset, the kernel function can be expressed as a matrix:\\
\begin{equation}
	K_{ij}=\ k\left(x_i,\ x_j\right) 
\end{equation}

QSVM integrates the architecture of quantum computing with classical computing to obtain the final results. The corresponding structural diagram is shown in Figure 1 below:\\

\begin{figure}[ht]
    \centering
    \includegraphics[width=0.8\textwidth]{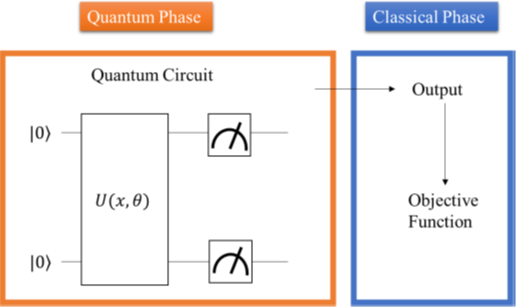}
    \caption{The corresponding structural diagram.}
    \label{fig:fig1}
\end{figure}

The solution process is explained by the flowchart in Figure 2:\\
\begin{figure}[ht]
    \centering
    \includegraphics[width=0.5\linewidth]{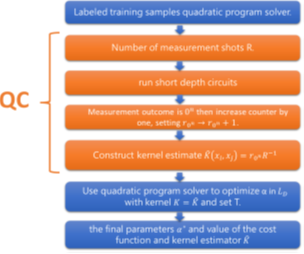}
    \caption{Flowchart of QSVM solution process.}
    \label{fig:flowchart}
\end{figure}

As shown in Figure \ref{fig:fig1}, this is the actual image.

\begin{enumerate}
    \item \textbf{Input Labeled Training Samples} \\
    The process begins with providing labeled training samples, which serve as the foundational data for subsequent model training.\\

    \item \textbf{Quadratic Program Solver} \\
    The training samples are fed into a quadratic program solver, which optimizes the model parameters.\\

    \item \textbf{Set Number of Measurements} \\
    Entering the quantum computing phase, the number of measurement shots \( R \) is determined. This represents the number of times the quantum circuit is executed to collect statistical data.\\

    \item \textbf{Run Short-Depth Circuits} \\
    Next, short-depth circuits are executed, which is the core step of quantum computing. Short-depth circuits are designed to minimize quantum noise and improve the reliability of computational results.\\

    \item \textbf{Process Measurement Outcomes} \\
    The quantum measurement outcomes are processed. If the measurement outcome is 0, the counter is incremented by one, and the variable \( r_0^n \) is updated as \( r_0^n + 1 \).\\

    \item \textbf{Construct Kernel Estimate} \\
    After completing the measurements, the kernel estimate is calculated using the following formula:

    \begin{equation}
    \hat{K}\left(x_i, x_j\right) = \frac{r_0^n}{R}
    \end{equation}

    Where \( r_0^n \) is the number of times the measurement outcome is 0, \( R \) is the total number of measurements, and \( \hat{K}\left(x_i, x_j\right) \) is the estimated kernel function value.\\

    \item \textbf{Optimize Parameters} \\
    The kernel estimate \(\hat{K}\) is sent back to the quadratic program solver to optimize the parameters \(\alpha\) in the target function \(L_D\), and the related variable \(T\) is set to determine the optimal model parameters.\\

    \item \textbf{Output Final Results} \\
    Finally, the process outputs the optimal parameters \(\alpha^\ast\), the cost function value, and the kernel estimate \(\hat{K}\), marking the completion of model training.\\

    \begin{equation}
    L(\alpha) = \sum_{i=1}^{t} \alpha_i - \sum_{i,j=1}^{t} y_i y_j \alpha_i \alpha_j K(x_i, y_i)
    \end{equation}
    \begin{equation}
    \text{s.t.} \quad \sum_{i=1}^{t} \alpha_i y_i = 0, \quad \alpha_i \geq 0
    \end{equation}
    
    Here, \(y_i\) indicates whether the labels belong to the same class. If \(y_i = y_j\) (same class), then \(y_i y_j = 1\); if \(y_i \neq y_j\) (different classes), then \(y_i y_j = -1\). The constraint in equation (5) ensures a balance between positive and negative classes. For non-support vectors, \(\alpha_i = 0\). For support vectors, \(\alpha_i > 0\), with its specific value depending on the distance of the data point \(x_i\) from the classification boundary and the kernel function \(K\left(x_i, x_j\right)\).\\

\end{enumerate}
In the above process, steps 3 to 6 are executed by the quantum computer, with the design of logic gates on the Qiskit platform illustrated in Figures 3 and 4. The remaining steps are performed using classical computing.

\begin{figure}[ht]
    \centering
    \includegraphics[width=0.5\linewidth]{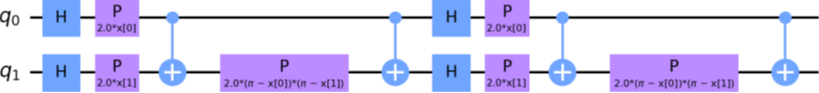}
    \caption{Two-dimensional quantum support vector machine circuit diagram implemented on the Qiskit platform.}
    \label{fig:qsvm-2d}
\end{figure}

\begin{figure}[ht]
    \centering
    \includegraphics[width=0.5\linewidth]{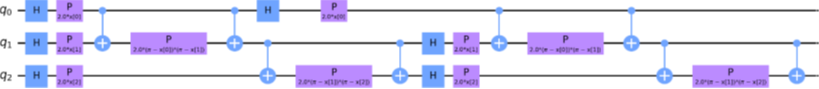}
    \caption{Three-dimensional quantum support vector machine circuit diagram implemented on the Qiskit platform.}
    \label{fig:qsvm-3d}
\end{figure}

\section{Introduction to VQC Methodology}
The VQC adopts a variational approach, constructing a classifier through a parameterized quantum circuit \(W(\theta)\) and a feature map \(U_{\mathrm{\Phi}(x)}\). The VQC circuit (as shown in Figure 5) is divided into three parts. The first part is data encoding, which transforms classical data sets into a Hilbert feature space using a nonlinear function. The second part is the variational optimization circuit, which utilizes the variational optimization circuit to find the optimal value \(\theta\). The final part is measurement. The circuit output is:
\begin{equation}
\mid\psi(x,\theta)\rangle = W(\theta)U_{\mathrm{\Phi}(x)}\mid{0\rangle}^{\otimes n}
\end{equation}

The classification probability is determined by the measurement operator \(M_y\), such as measuring the expectation value \(\langle Z \rangle\) of the first qubit. The loss function \(L(\theta)\) is defined as the expectation of prediction errors:
\begin{equation}
L(\theta) = \sum_{i=1}^{t} P_r(\widetilde{m}(x_i) \neq y_i \mid x_i \in T)
\end{equation}
where the error probability is approximated as:
\begin{equation}
P_r(m(x_i) \neq y_i \mid x_i \in T) \approx \mathrm{sig}\left(\sqrt{R} \cdot \frac{2^{-y_i b} p_y}{\sqrt{2 p_y (1-p_y)}}\right)
\end{equation}

Here, \( p_y = \frac{r_y}{R} \), \(\mathrm{sig}(x) = \frac{1}{1+e^{-x}} \), and \( R \) is the number of measurements. The parameters \(\theta\) are optimized via gradient descent: \(\theta_{t+1} = \theta_t - \eta \nabla L(\theta)\), where \(\hat{m}(x_i)\) is the model's predicted label for data point \(x_i\), typically a classification result (e.g., +1 or -1 in binary classification). \(\hat{m}(x_i) \neq y_i\) indicates a prediction error, and the error probability \(\Pr(\hat{m}(x_i) \neq y_i \mid x_i \in T)\) measures the model's performance on the test set \(T\). Here, \(\eta\) is the learning rate, and the gradient \(\nabla L(\theta)\) can be computed using the parameter-shift rule. The goal of the entire training process is to minimize the value of \(L(\theta)\).

\begin{figure}[ht]
    \centering
    \includegraphics[width=0.5\linewidth]{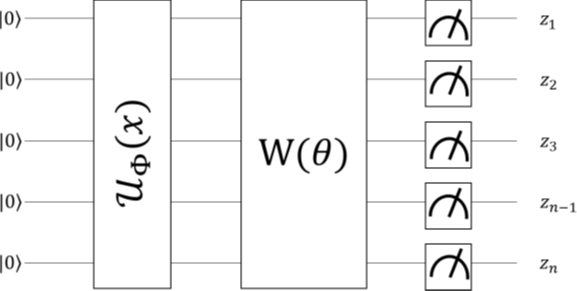}
    \caption{VQC circuit diagram.}
    \label{fig:vqc-circuit}
\end{figure}

The execution process is as follows:
\begin{enumerate}
    \item \textbf{Initialization:} \\
    Set the initial quantum circuit parameters \(\theta_0\), select the feature mapping circuit \(U_{\mathrm{\Phi}(x)}\), and the parameterized circuit \(W(\theta)\). Initialize the number of measurements \(R\) and the learning rate \(\eta\).

    \item \textbf{Forward Propagation:} \\
    For each training data point \(x_i\), run the quantum circuit \(\{W(\theta)U_{\mathrm{\Phi}}(x_i)\mid0\rangle\}^{\otimes n}\), measure the output, and calculate the empirical distribution \(\hat{p}_y(x_i) = \frac{r_y}{R}\), where \(r_y\) represents the number of times the measurement result corresponds to class \(y\) in \(R\) quantum measurements.

    \item \textbf{Loss Calculation:} \\
    Based on the measurement results, estimate the error probability \(L(\theta)\), and adjust the parameters \(\theta\) iteratively to achieve a lower error rate.

    \item \textbf{Gradient Calculation:} \\
    For each parameter \(\theta_j\), use the parameter-shift rule:
    \begin{equation}
    \frac{\partial L}{\partial \theta_j} = \frac{L(\theta_j + \pi/2) - L(\theta_j - \pi/2)}{2}
    \end{equation}

    \item \textbf{Parameter Update:} \\
    Apply gradient descent:
    \begin{equation}
    \theta_{t+1} = \theta_t - \eta \nabla L(\theta_t)
    \end{equation}
    Repeat Steps 2-5 until convergence.

    \item \textbf{Prediction:} \\
    For test data \(x\), run the optimized circuit \(W(\theta^\ast)U_{\mathrm{\Phi}(x)}\mid0\rangle^{\otimes n}\), and assign the label \(y = \arg\max_y \hat{p}_y(x)\) based on the measurement distribution \(\hat{p}_y(x)\).

\end{enumerate}

\begin{figure}[ht]
    \centering
    \includegraphics[width=0.5\linewidth]{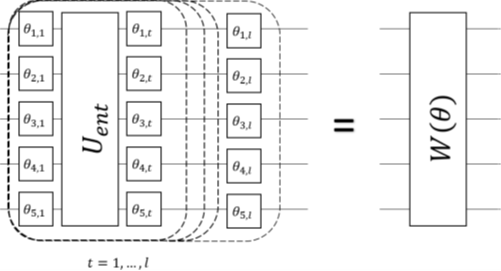}
    \caption{Entanglement circuit with repeated layers for VQC.}
    \label{fig:entanglement-circuit}
\end{figure}

As shown in Figure 6 above, we applied \(l\) repeated entanglement circuits and rotated the logic gates of all single qubits to achieve the purpose of quantum computing. The circuit for \(U_{\mathrm{ent}}\), as shown in Figure 7, consists of five SWAP two-qubit circuits.

\begin{figure}[ht]
    \centering
    \includegraphics[width=0.5\linewidth]{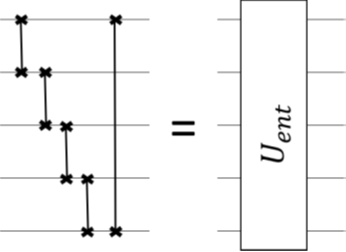}
    \caption{SWAP two-qubit circuit for $U_{\mathrm{ent}}$.}
    \label{fig:swap-circuit}
\end{figure}

\section{Result and Discussion}
Several biostatistical metrics serve as critical indicators for evaluating the performance of predicting cell surface epitope characteristics, namely accuracy, AUC, and MCC, in that order. Among them, accuracy (ACC) reflects the overall proportion of correct predictions made by the method. A higher ACC value indicates better performance in distinguishing epitopes from non-epitopes. AUC, defined based on the ROC curve, ranges from 0 to 1, where 1 signifies a perfect classifier, and 0.5 indicates performance equivalent to random guessing. MCC, a more comprehensive metric, accounts for true positives, false positives, true negatives, and false negatives. Its value ranges from -1 to 1, with 1 representing perfect prediction, 0 indicating random prediction, and -1 denoting completely incorrect prediction.\\

\begin{table}[ht]
\centering
\setlength{\tabcolsep}{10pt} 
\renewcommand{\arraystretch}{4} 
\begin{tabular}{llccc}
\toprule
\textbf{Research paper} & \textbf{Year} & \textbf{ACC} & \textbf{AUC} & \textbf{MCC} \\
\midrule
\multirow{2}{*}{\parbox{5.5cm}{ABCPred: A Server for Predicting B-cell Epitopes Using Artificial Neural Networks}} & 2005 & 66\% & 0.72 & 0.2 \\
\multirow{2}{*}{\parbox{5.5cm}{Prediction of continuous B-cell epitopes in an antigen using recurrent neural network}} & 2006 & 65\% & 0.64 & \multirow{2}{*}{X} \\
\multirow{2}{*}{\parbox{5.5cm}{Improved method for predicting linear B-cell epitopes}} & 2006 & 65\% & 0.70 & \\
\multirow{2}{*}{\parbox{5.5cm}{PEPITO: improved discontinuous B-cell epitope prediction using multiple distance thresholds and half sphere exposure}} & 2008 & 70\% & 0.75 & \multirow{2}{*}{X} \\
\multirow{2}{*}{\parbox{5.5cm}{SVMtriP: a method to predict antigenic epitopes using support vector machine to integrate tri-peptide similarity and propensity}} & 2012 & 70\%-75\% & 0.74 & 0.4 \\
\multirow{2}{*}{\parbox{5.5cm}{BepiPred-2.0: improving sequence-based B-cell epitope prediction using conformational epitopes}} & 2017 & 68\%-72\% & 0.74 & \multirow{2}{*}{X} \\
\multirow{2}{*}{\parbox{5.5cm}{Linear B-Cell Epitope Design: A Performance Review of Methods Available via Command-Line Interface}} & 2021 & 65\%-70\% & 0.68-0.72 & 0.3-0.4 \\
\multirow{2}{*}{\parbox{5.5cm}{A Structure-Based B-cell Epitope Prediction Model Through Combining Local and Global Features}} & 2022 & 75\% & 0.76 & 0.45 \\
\multirow{2}{*}{\parbox{5.5cm}{B cell epitope prediction by capturing the spatial clustering property of the epitopes using graph attention network}} & 2024 & 78\% & 0.8 & 0.5 \\
\multirow{2}{*}{\parbox{5.5cm}{QSVM}} & 2025 & 70\% & 0.71 & 0.42 \\
\multirow{2}{*}{\parbox{5.5cm}{VQC}} & 2025 & 73\% & 0.703 & 0.148 \\
\bottomrule
\end{tabular}
\caption{Comparison of B-cell epitope prediction methods.}
\label{tab:bcell_prediction}
\end{table}

When comparing QSVM and VQC, two quantum computing-based methods for B-cell epitope prediction, with other classical computing methods, several notable characteristics and advantages emerge. First, in terms of accuracy (ACC), VQC achieves a leading accuracy of 73\%, surpassing most classical methods and only trailing the 2024 method (78\%), while QSVM performs comparably to most classical methods (70\%). This highlights the potential of quantum computing to enhance prediction accuracy.\\

Regarding the area under the ROC curve (AUC), QSVM (0.71) and VQC (0.70) exhibit moderate performance, slightly below certain classical methods such as PEPITO (0.75) and the 2024 method (0.80), but still within a reasonable range. This indicates the stability of quantum computing methods in distinguishing epitopes from non-epitopes.\\

In terms of the Matthews Correlation Coefficient (MCC), QSVM’s score of 0.42 stands out, outperforming many classical methods, such as the 2021 method (0.3–0.4) and ABCpred (0.2), demonstrating its advantage in balancing positive and negative samples. However, VQC’s MCC of only 0.148 may reflect challenges in handling sample balance.\\

Overall, QSVM and VQC demonstrate the potential of quantum computing in B-cell epitope prediction, particularly in terms of accuracy and MCC. Nevertheless, there remains room for improvement in AUC and MCC. The advancements of quantum computing methods provide new directions for future research. With further algorithm optimization and improvements in quantum hardware, significant progress in prediction accuracy and reliability is anticipated. These results suggest that quantum computing holds broad prospects for applications in biology.

\section{Conclusion}
This study explores the application of quantum computing in B-cell epitope prediction, demonstrating the potential of quantum machine learning in bioinformatics through the use of Quantum Support Vector Machines (QSVM) and Variational Quantum Classifiers (VQC). B-cell epitope prediction is a core task in immunological research, holding significant implications for vaccine design, diagnostics, and therapeutic development. This research not only highlights the prospects of quantum algorithms in enhancing prediction accuracy and efficiency but also identifies current challenges and future directions for development.\\

The study first reviews the historical evolution of B-cell epitope prediction methods, starting from BcePred in 2004, which relied on physicochemical properties, to subsequent machine learning approaches such as SVMTriP and BepiPred-2.0. These traditional methods have progressively improved prediction accuracy, reaching a peak of 55.59\% (e.g., the Consensus ALL method in 2021). However, as the scale and complexity of biological datasets increase, the computational complexity of traditional methods often becomes inadequate, prompting researchers to explore quantum computing as a solution. Quantum computing leverages the properties of quantum entanglement and superposition, theoretically enabling more efficient handling of such computationally intensive problems. This study applies QSVM and VQC for B-cell epitope prediction, achieving accuracies of 70\% and 73\%, respectively. These results are competitive with traditional methods and, in some aspects, outperform them, demonstrating the potential of quantum algorithms in bioinformatics applications.\\

A major contribution of this study lies in the detailed implementation of QSVM and VQC, optimized specifically for the B-cell epitope prediction problem. The QSVM, grounded in the principles of classical support vector machines, employs a quantum kernel function to map data into a high-dimensional feature space and constructs a hyperplane to distinguish epitopes from non-epitopes. The optimization problem is defined as \(L\left(\alpha\right)=\sum_{i=1}^{t}\alpha_i-\sum_{i,j=1}^{t}{y_iy_j\alpha_i\alpha_jK\left(x_i,x_j\right)}\), subject to the constraints \(\sum_{i=1}^{t}{\alpha_iy_i}=0\) and \(\alpha_i\geq0\), solved using a quadratic programming solver. The kernel function \(K\left(x_i,x_j\right)\) is estimated by a quantum circuit, leveraging quantum computing’s advantage in computing inner products in high-dimensional spaces, theoretically reducing computational overhead. On the other hand, the VQC adopts a variational approach, constructing a classifier using a parameterized quantum circuit \(W\left(\theta\right)\) and a feature map \(U_{\Phi\left(x\right)}\). The training process involves computing the loss function \(L\left(\theta\right)\) through quantum measurements and iteratively updating the parameters \(\theta_{t+1}=\theta_t-\eta\nabla L\left(\theta_t\right)\). The empirical probability distribution \(\widehat{p_y}\left(x_i\right)=\frac{r_y}{R}\) is derived from \(R\) measurements and used to assign class labels, demonstrating the practicality of quantum measurements in classification tasks.\\

The performance comparison between QSVM and VQC, as well as their benchmarking against traditional methods, provides valuable insights. VQC achieved a slightly higher accuracy (73\%) compared to QSVM (70\%), indicating its effectiveness in capturing complex patterns of B-cell epitopes. However, in terms of the Matthews Correlation Coefficient (MCC), QSVM scored 0.42, significantly outperforming VQC’s 0.148, suggesting that QSVM is superior in balancing predictions between positive and negative classes. This discrepancy may be attributed to VQC’s sensitivity to class imbalance, with its lower MCC potentially stemming from the empirical probability estimation process \(\widehat{p_y}\left(x_i\right)\). Comparisons with traditional methods reveal that both quantum approaches perform comparably to the 2022 structural model (accuracy 75\%, MCC 0.45) and the 2024 graph attention network model (accuracy 78\%, MCC 0.5). Notably, the area under the ROC curve (AUC) for QSVM and VQC was 0.71 and 0.703, respectively, slightly lower than traditional methods (e.g., 0.8 for the 2024 model), indicating room for improvement in their ability to distinguish epitopes from non-epitopes.\\

The error probability analysis of VQC further elucidates the challenges in quantum classification, as expressed by the formula \(P_r\left(\hat{m}\left(x_i\right)\neq y_i\mid x_i\in T\right)\approx\mathrm{sig}\left(\sqrt{R}\cdot\frac{2^{-y_ib}p_y}{\sqrt{2p_y\left(1-p_y\right)}}\right)\). This formula indicates that prediction errors are closely tied to the number of measurements \(R\), the bias term \(b\), and the prior probability \(p_y\). Increasing \(R\) can reduce statistical noise in probability estimation but also elevates computational costs, presenting a trade-off that must be balanced in practical applications. Furthermore, VQC’s low MCC highlights its difficulties in handling imbalanced datasets, a common issue in bioinformatics where non-epitopes significantly outnumber epitopes. Future research could explore cost-sensitive learning or oversampling techniques within a quantum framework to address this challenge.\\

Despite the encouraging results, several challenges remain. Even on platforms like Qiskit, the computational resource demands of quantum circuits are substantial, and current quantum hardware limitations impact the scalability of QSVM and VQC on large-scale datasets. Additionally, the optimization of variational parameters in VQC may encounter the "barren plateau" problem of vanishing gradients. Although this study mitigates this issue by computing gradients using parameter-shift rules, more robust optimization techniques, such as quantum-aware optimizers, could further enhance performance. The AUC and MCC metrics indicate that while quantum methods are competitive, they have not yet surpassed the best traditional methods in all aspects, suggesting that hybrid quantum-classical approaches may be a promising direction for future research.\\

Overall, this study demonstrates the potential of QSVM and VQC in B-cell epitope prediction, achieving accuracies of 70\% and 73\%, respectively, and highlighting the advantages of quantum computing in handling complex biological data. The higher MCC of QSVM (0.42) underscores its robustness in classification tasks, while VQC’s superior accuracy (73\%) reflects its capability in pattern recognition. However, challenges such as class imbalance, computational costs, and optimization difficulties must be addressed to fully realize the potential of quantum machine learning in bioinformatics. Future research should focus on improving quantum hardware, developing hybrid algorithms, and exploring quantum-enhanced feature selection techniques to further boost prediction performance. As quantum computing technology matures, its applications in epitope prediction and other bioinformatics problems hold the promise of revolutionizing the field, offering more accurate and efficient solutions for immunology and related domains, and opening new avenues for research and application.\\

\bibliographystyle{unsrtnat}
\bibliography{references}

\end{document}